# Room Temperature Magneto-dielectric coupling in the CaMnO$_3$ modified NBT lead-free ceramics

Koyal Suman Samantaray[1], Ruhul Amin[1], Saniya Ayaz[1], A. K. Pathak[2], Christopher Hanley[2], A. Mekki[3,4], K. Harrabi[3,4] Somaditya Sen[1*]

[1]Department of Physics, Indian Institute of Technology Indore, Indore, 453552, India

[2]Department of Physics, SUNY, Buffalo State, NY, USA

[3]Department of Physics, King Fahd University of Petroleum & Minerals Dhahran, 31261, Saudi Arabia

[4]Center for Advanced Material, King Fahd University of Petroleum & Minerals, Dhahran, 31261, Saudi Arabia

*Corresponding Author: sens@iiti.ac.in

Abstract:

The sol-gel prepared (1-x) Na$_{0.5}$Bi$_{0.5}$TiO$_3$- (x) CaMnO$_3$ (x=0, 0.03, 0.06, 0.12) compositions show a Rhombohedral (*R3c*) phase for x=0.06 while a mixed Rhombohedral (*R3c*) and orthorhombic (*Pnma*) phases for the x=0.12. The lattice volume consistently decreased with an increase in the CaMnO$_3$ content. The phase transition temperature (T$_c$) decreased with an increase in the CaMnO$_3$ compositions. The room temperature dielectric constant increased, and loss decreased for the x=0.03 composition due to a decrease in the oxygen vacancy and Bi loss confirmed by the valence state study (XPS). All the compositions show a variation of the room temperature dielectric property with an application of magnetic field confirming a magnetodielectric coupling. The x=0.06 composition shows the highest negative magnetodielectric constant (MD%) of 3.69 at 100kHz at an applied field of 5 kG.

Introduction:

The tuning of dielectric properties with magnetic fields is one of the most trending research problems in the current scenario. The presence of magnetoelectric (ME) coupling improves the functionality of the materials, making it one of the promising candidates for various device applications such as sensors, four-state logic in one device, spintronics, FeRAMs, MRAMs, etc. [1,2]. Such magnetoelectric coupling is best observed in the multiferroics that simultaneously show different ferroic orders such as ferroelectricity, ferromagnetism, antiferromagnetism, etc. [3,4]. The ferroelectric materials can be modified by introducing magnetism to them, which leads to a magnetoelectric coupling in the engineered materials [5,6]. However, the single-phase multiferroic materials are rare, as it requires the coexistence of two mutually exclusive phenomena, i.e., cation off-centering in ferroelectrics (which originates due to the $d^0$ orbitals) and the formation of magnetic moments (which develops due to the partially filled d or f orbitals) [7,8]. The magnetism can be introduced in a ferroelectric material due to the incorporation of (i) magnetic ions, (ii) self-defects such as cationic and oxygen vacancy, (iii) surface defects, and (iv) magnetic clusters, and (v) lattice strain [9–11]. Such effects develop the coupling of the magnetic and ferroelectric orders by efficiently controlling the magnetic domains by applying an electric field [12].

Among various lead-free materials, $Na_{0.5}Bi_{0.5}TiO_3$ (NBT) is one of the best choices for producing magnetoelectric coupling [13,14]. It shows a high dielectric constant ($\varepsilon_r \sim 800$), strong ferroelectric polarization ($P_r \sim 38$ $\mu C.cm^{-2}$), high curie temperature ($T_c \sim 320$ °C), and high piezoelectric coefficient ($d_{33} = 58$-$95$ pC/N) [15,16]. The report by Jain Ruth D.E. et al. discussed the presence of ferromagnetism in the NBT at a magnetic field lower than 800 Oe [17]. This phenomenon was related to the presence of Na-vacancy in the samples. The observance of a high ME coupling coefficient of 4.18mV/cm Oe observed at zero DC magnetic bias field was reported. Another report by Lin Ju et al.; shows a giant room temperature magneto-dielectric constant (MD%) of 9.48% at 1 kHz under H=8 kOe [18]. This was attributed to the ferromagnetism originating from the Na-vacancy at the NBT (1 0 0) surface. Apart from these self-defects, magnetism was also achieved by the incorporation of various elements like Fe, Mn, Ni, Co, and Cu and by forming solid solutions of NBT with Bi ($Ti_{1/2}Co_{1/2}$) $O_3$, $SrCoO_{3-\delta}$, $CoTiO_3$, and $MgCoO_{3-\delta}$, etc. [10,14,19–21]. Bulk $CaMnO_3$ (CMO) is an exciting material due to its room temperature antiferromagnetic and paraelectric properties [22]. But several reports have proved the possibility of room temperature multiferroicity in CMO by doing strain and chemical engineering [23]. One report by Dung et al.; reported the presence of ferromagnetism in the solid solution of (1-*x*) NBT-*x* CMO [24]. Their article has discussed the optical and magnetic properties of the series up to 9% substitution only. However, a complete discussion on the structure correlation, Morphotropic Phase Boundary (MPB), dielectric, and magneto-dielectric coupling of these series of materials were not provided. In the present work, a series of (1-*x*) NBT-*x*CMO *(x=0, 0.03, 0.06, 0.12)* compositions are critically analyzed from the structural/vibrational studies point of view using XRD and Raman studies, with support from a valence state study using XPS studies. The presence of an MPB region is also investigated in detail in this series of compositions. The effect of these factors on the dielectric properties was discussed. The shifting of phase transition temperature (Tc) obtained from the temperature-dependent dielectric study with the incorporation of CMO is also elaborately discussed. Further magnetism is explored, and the effect of the magnetic field on the dielectric properties is also investigated. The coupling of dielectric polarization with magnetic field and the possibility of magnetoelectricity at room temperature is detailed.

Methodology:

(1-x) $Na_{0.5}Bi_{0.5}TiO_3$-$xCaMnO_3$ (x=0, 0.03, 0.06, and 0.12) polycrystalline powders, were synthesized using modified sol-gel technique. Our previous work also mentioned a similar synthesis route [25]. For this process sodium nitrate (purity 99.9%) for Na, bismuth nitrate pentahydrate (purity 98%) for Bi, Calcium nitrate (purity 99.9%) for Ca, dihydroxy bis (ammonium lactate) titanium (IV), 50% w/w aqua solution (purity 99.9%) for Ti, Manganese nitrate (purity 99.9%) for Mn were used as precursors. All the above-mentioned precursor materials were purchased from Alfa Aesar. The solutions of individual precursors were mixed to get a clear solution which was further stirred continuously for an hour. The citric acid and ethylene glycol were mixed in the molar ratio of 1:1 and continuously stirred for another hour. The final mixture was stirred and maintained at ~80°C until a clear gel started to form. The burnt gel powders were ground carefully and heated at 450ºC for 12 h for decarburization and denitrification and further annealed at 750 °C for phase formation. The phase was verified from x-ray diffraction (XRD) using an x-ray diffractometer (Bruker D2-Phaser). All the physical properties reported in this work will be on these sintered samples prepared at 1100°C-1130°C for 3hr.

The XRD pattern of all the samples was refined using Rietveld refinement. The refinement was done using the FullProf suite software considering pseudo-Voigt peak shape for all the samples. The chemical composition and the valence states of the elements present in the prepared samples were studied using the X-Ray Photoelectron Spectroscopy (XPS) experiment performed using the Thermo-Scientific Escalab 250 Xi XPS Spectrometer (Al-Kα x-rays) having an energy resolution of ~ 0.5 eV. The spectra were deconvoluted using XPSPEAK41 software. The XPS spectra were initially calibrated using the adventitious C1$s$ peak at a binding energy of 284.8 eV. The background was extracted correctly using the Tougaard function. A combined Gaussian-Lorentzian peak shape was used for all the peaks to get quality fitting. Room temperature phonon modes were studied from Raman spectroscopy using Horiba-made LabRAM HR Evolution Raman spectrometer (spectral resolution 0.9 cm$^{-1}$) having He-Ne LASER of wavelength 632.8 nm. The electrical characterization for all the samples was done on the sintered pellets of the final phase polycrystals. The silver electrodes were prepared on both sides of the pellets for dielectric measurement using silver paste. The pellets were after that annealed for proper adhesion of the Ag to the pellets at 540ºC for 15 minutes. The morphological studies of the sintered pellet samples were done using a Supra55 Zeiss Field Emission Scanning Electron Microscope (FE-SEM). The dielectric measurement was done using a broadband dielectric spectrometer using Newton's 4$^{th}$ Ltd. phase-sensitive multimeter having signal strength 1V$_{rms}$ in the temperature range 50ºC to 450ºC and frequency range 1Hz-1MHz. The magnetic properties were performed in a Physical Property Measurement System (PPMS, Quantum Design) using a vibrating sample magnetometer (VSM) option. The magnetodielectric measurement was carried out using the electromagnet connected to a current source and digital gauss meter (SES Instruments Pvt. Ltd.). The corresponding room temperature dielectric was measured using the Newton's 4$^{th}$ Ltd. phase-sensitive multimeter having signal strength 1V$_{rms}$.

Results and Discussions:

X-Ray Diffraction

The phase of the (1-x) $Na_{0.5}Bi_{0.5}TiO_3$ - (x)$CaMnO_3$ solid solution with x≤0.6 was observed to be in the Rhombohedral (*R3c*) phase. For higher values of substitution. i.e., for x=0.12 a mixed phase of Rhombohedral (*R3c*) and orthorhombic (*Pnma*) structures were observed [Fig.1(a)]. $CaMnO_3$ is known to be in the orthorhombic (*Pnma*) phase at room temperature. The mixed-phase is due to the higher concentration of $CaMnO_3$ [26]. The obtained XRD pattern was refined using Rietveld refinement by taking pseudo-Voigt peak profiles to study the structures in detail further. The goodness of fit ($\chi^2$) was within acceptable limits [Fig.1(b)-(e)]. The Rietveld analysis shows a presence of 25% *Pnma* and 75% *R3c* phase for x=0.12. The lattice volume was observed to decrease with an increase in the $CaMnO_3$ content. The Shannon radii of $Ca^{2+}$(XII) ~1.34 Å ions is smaller than that of the $Na^+$(XII) ~ 1.39 Å and $Bi^{3+}$(XII) ~1.36 Å ions. Also, the $Mn^{3+}$ (VI-LS) ~0.58 Å, and $Mn^{4+}$~0.53 Å are smaller in comparison to the $Ti^{4+}$ (VI) ~0.605 Å, while $Mn^{2+}$ (VI-LS) ~0.67 Å, is larger than $Ti^{4+}$ (VI) but is comparable to $Ti^{3+}$ (VI) ~0.67 Å [27]. One expects an $Mn^{4+}$ ion to substitute a $Ti^{4+}$ ion. Hence, this leads to a decrease in the lattice volume with substitution. The lattice strain was calculated using Scherrer's equation from the FWHM of the XRD peaks [28]. The strain was observed to decrease for x=0.03 and thereafter continuously increase for the other two compositions. To understand such a trend in the lattice strain, the tilt angle of the octahedra was calculated from the atomic positions [29]. The octahedral tilt angle was observed to be responsible for such lattice strain variation [25]. The lattice parameters (a=b), after experiencing an increase for x=0.03, continuously decreased for the other two compositions [Fig. 1(f)]. On the other hand, the "c" parameter shows a sporadic up and down variation with an increase in $CaMnO_3$ composition [Fig. 2(a)]. To study such variation in the lattice parameters, a detailed bond length and bond angle analysis is done.

In the $BO_6$ octahedra, there are two sets of B-O bond lengths. One set is larger than another one [25]. The short B-O bond lengths vary similarly to the variation in the "c" lattice parameter [Fig.2 (b)] The long B-O bond lengths vary exactly opposite to the shorter ones. It can be concluded that the "c" lattice parameter is affected by the B-O bond lengths variation. The O-B-O bonds associated with the B-O bonds also show a similar trend with the composition [Fig.2 (d)]. There are four types of A-O bonds present in the NBT-based rhombohedral structure, denoted as $A_1$-O, $A_2$-O, $A_3$-O, and $A_4$-O. The $A_1$-O and the $A_2$-O bond lengths increased for x=0.03 and thereafter continuously decreased for the other two compositions while an exactly opposite trend was observed for the $A_3$-O and $A_4$-O bonds [Fig.2(c)]. A similar variation was observed for the associated A-O-A bond angles [Fig.2(e)]. The variation in the displacement of B-site atoms with composition shows a similar trend as the variation in the B-O bond lengths [Fig.2 (f)]. The off-centering of the O-atoms increased for the x=0.03 and further decreased continuously for the other compositions. Such variation in the off-centering of O-atoms is responsible for the A-O bond variations, consequently affecting the lattice strain.

The bond lengths and the lattice strain are affected by various cationic and anionic defects present in the lattice. NBT is known to show cationic (Na/Bi) vacancies due to the high-temperature sintering of the ceramics leading to Na and Bi loss. This leads to a proportionate oxygen loss creating oxygen vacancies [30]. To confirm such losses from the lattice, a detailed analysis of the XPS spectra is reported in the following section.

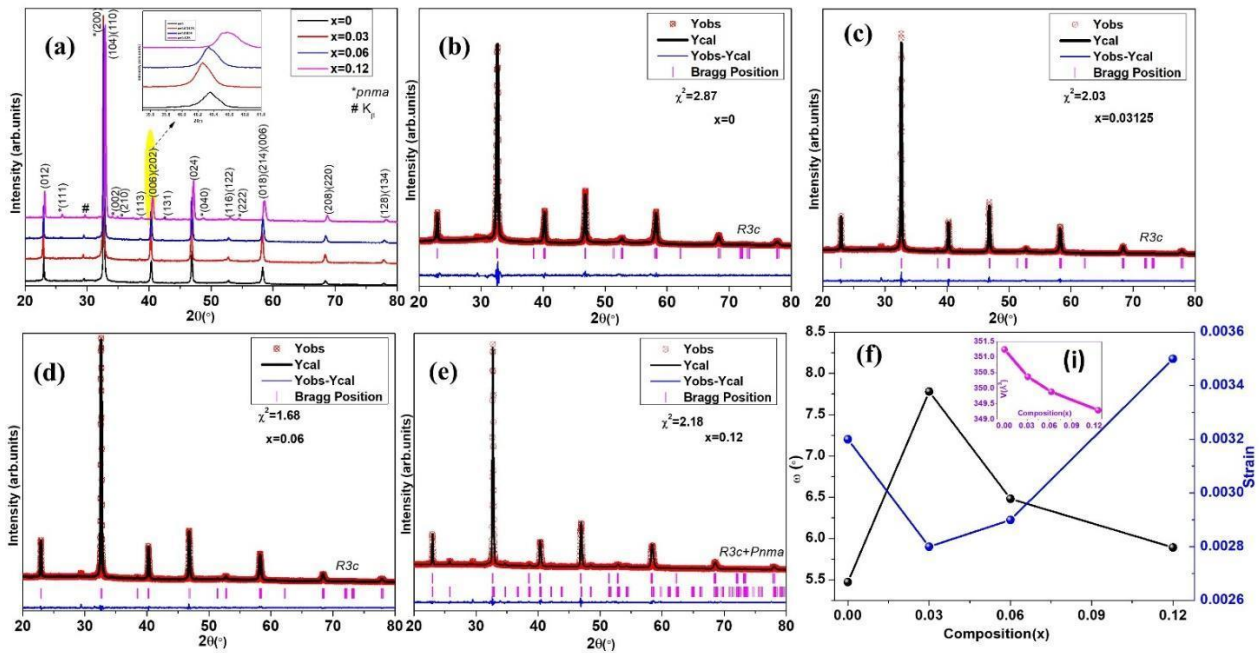

**Figure 1** (a) XRD pattern of (1-x) $Na_{0.5}Bi_{0.5}TiO_3$-xCaMnO$_{3-\delta}$ (x=0, 0.03, 0.06, and 0.12), Rietveld plot of (b) x=0 (c) x=0.03 (d) x=0.06 (e) x=0.12 (f) tilt angle and strain with composition; inset shows variation of volume with composition

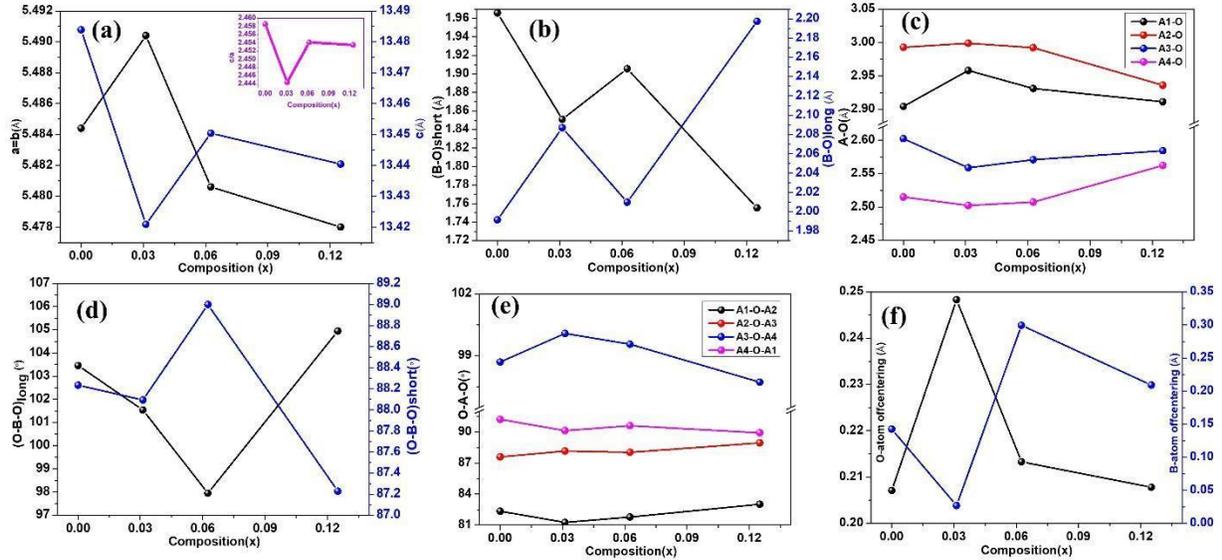

**Figure 2** (a) Variation of lattice parameters with composition (b) Variation of B-O bond lengths with composition (c) Variation of A-O bond lengths with composition (d) Variation of O-B-O bond angles with composition (e) Variation of O-A-O bond angles with composition (f) Variation of off-centering of the O-atom and B-atoms

XPS Analysis:

i. O 1$s$ core-level spectra:

The O-1$s$ spectra consisted of three peaks corresponding to the binding energy of lattice oxygen ($O_L$) at 529–530 eV, oxygen vacancy ($O_V$) at 530–532 eV, and adsorbed oxygen ($O_A$) at 533–534 eV [31,32]. The $O_L$ and $O_V$ are the ones that contribute to the lattice. The numerical estimation shows the $O_V$ fraction ($O_V/(O_V + O_L)$) decreased continuously for x=0 to x=0.06 but thereafter increased x=0.12 [Fig.3(a)-(d)]. Hence, the incorporation of CaMnO$_3$ seems to reduce the Ov in general. The least value of Ov was recorded for x=0.06. The creation of $O_V$ may be triggered by probable cation-loss in the compositions. Hence, a valence state study of the individual cations becomes significant. However, losses in lattice oxygen associated with cation loss imply probable distortions in the lattice which is expected to affect the lattice parameters and bond strengths.

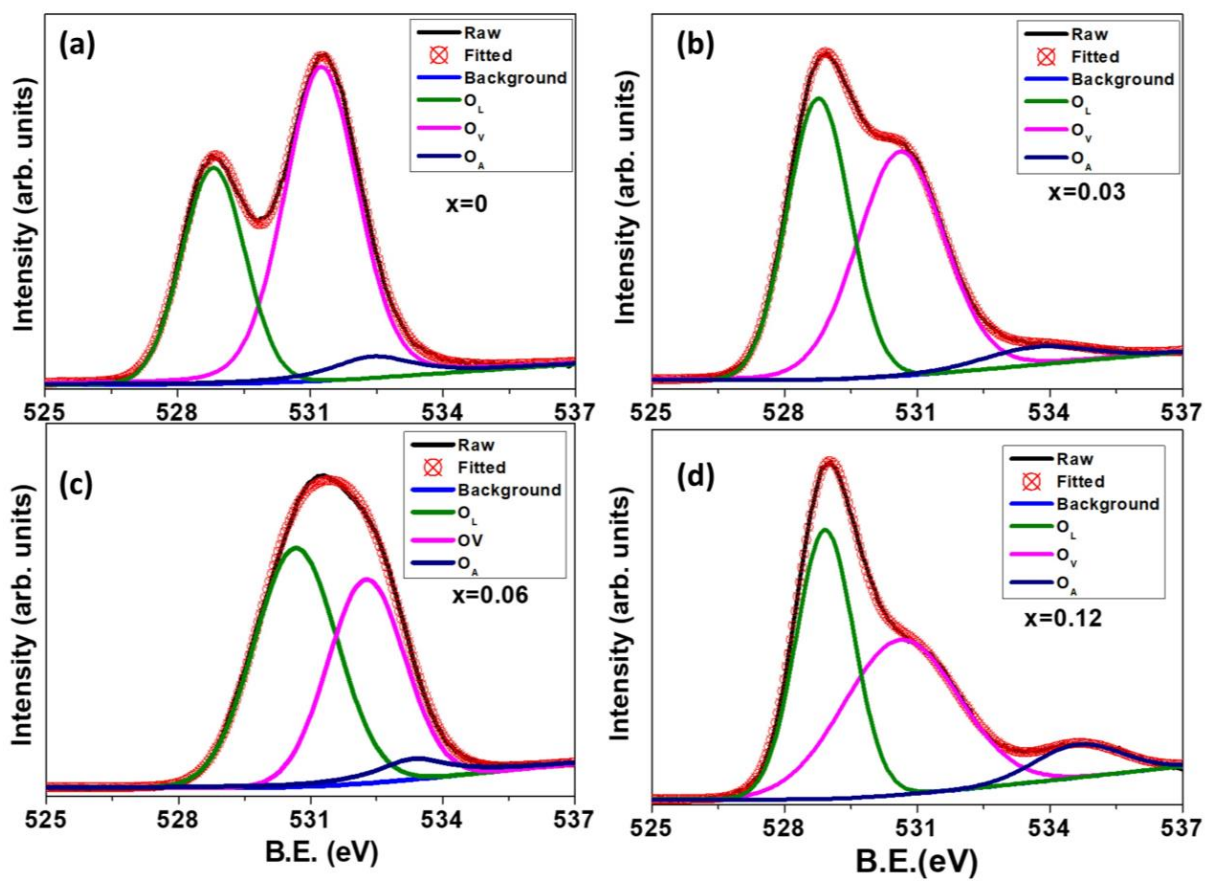

**Figure 3** Deconvoluted O-1s XPS pattern of (a) x=0, (b) x=0.03, (c) x=0.06, and (d) x=0.12 compositions

ii. Na 1$s$ core-level spectra:

The core Na 1$s$ peak at 1072.62 eV confirmed the Na$^+$ state for all the samples [Fig. 4(a)-(d)] [33]. Extra peaks observed in this region indicate the presence of Ti-LMM Auger transitions and will be discussed in the following section. The Ti-LMM Auger peaks are observed to be in the proximity of the Na 1$s$ peak for all the samples [34]. The binding energy of Na 1s peak continuously decreased from 1072.62 eV to 1071.44 eV with an increase in the CaMnO$_3$ composition. This indicates the weakening of Na-O bonds, which originated due to the volatilization of Na at high-temperature sintering [35].

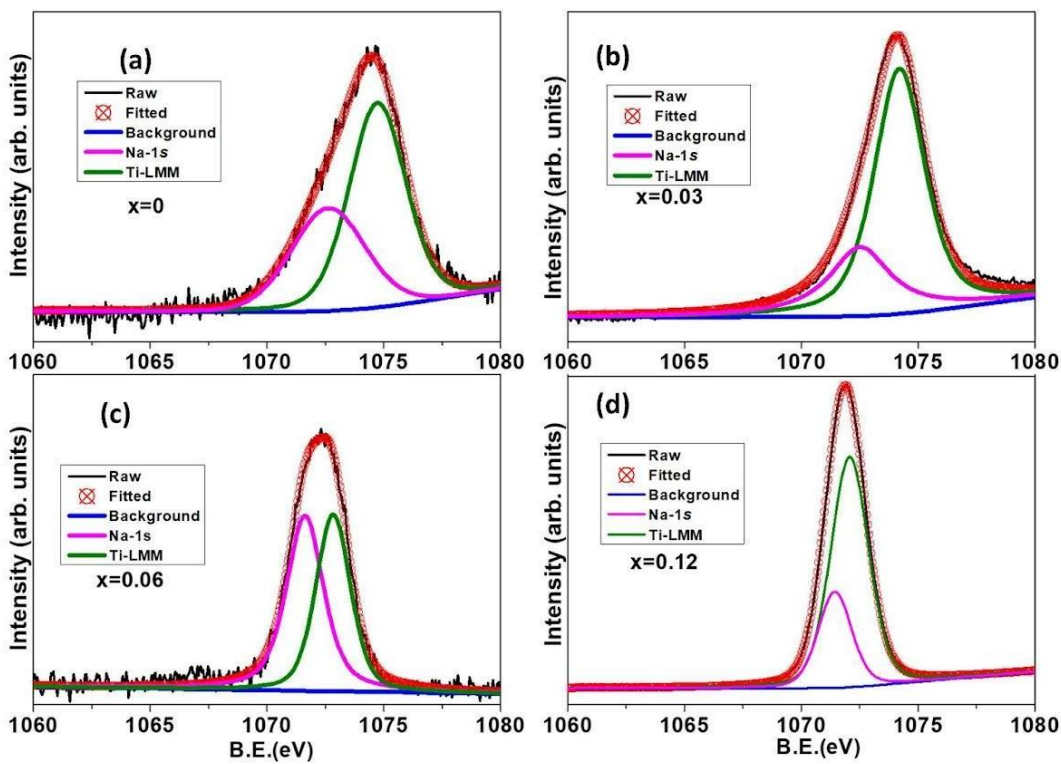

**Figure 4** Deconvoluted Na-1s XPS pattern of (a) x=0, (b) x=0.03, (c) x=0.06, and (d) x=0.12 compositions

iii. Bi 4$f$ core-level spectra:

Bi was observed to be in the $Bi^{3+}$ state. Along with the $Bi^{3+}$ states, Bi loss was observed for all the samples. However, the ratio of $Bi^{3+}$ to Bi-loss varied with the composition [Fig.5 (a)-(d)]. The $Bi^{3+}$ 4$f_{7/2}$ and $Bi^{3+}$ 4$f_{5/2}$ peak binding energy continuously decreased from (159.15 eV, 165.35 eV) for x=0 to (157.17 eV, 162.48 eV) for x=0.12 [36]. The spin-orbit splitting energy for $Bi^{3+}$4$f_{7/2}$ and $Bi^{3+}$4$f_{5/2}$ was observed to continuously decrease from 6.20 eV for x=0 to 5.31 eV for x=0.12. This decrease in B.E is due to the weakening of the Bi-O bonding due to Ca and Mn incorporation. The Bi-loss peaks were observed at (161.57 eV, 167.24 eV) for x=0, (162.11 eV, 166.98 eV) for x=0.03, (159.38 eV, 1164.70 eV) for x=0.06 and finally at (158.87 eV, 164.17 eV) for x=0.12 [37,38]. The spin-orbit splitting energy for Bi-loss peaks was observed to be decreased from 5.67 eV for x=0 to 4.98 eV for x=0.03 and then continuously increased to 5.30 eV for x=0.12 composition. The fraction of Bi-loss to $Bi^{3+}$ was estimated to decrease from 0.50 for x=0 to 0.37 for x=0.03 and further continuously increased to 0.89 for x=0.12. The variation of the spin-orbit splitting energy, binding energy, and the Bi-loss to $Bi^{3+}$ was observed to affect the A-O bond strength similarly, i.e., decreased for the x=0.03 composition and further continuously increased.

The Bi satellite peak detected in the Ti main peak region at 465.86 eV was observed to continuously shifted to a lower B.E. of 465.44 eV for x=0.12 [17].

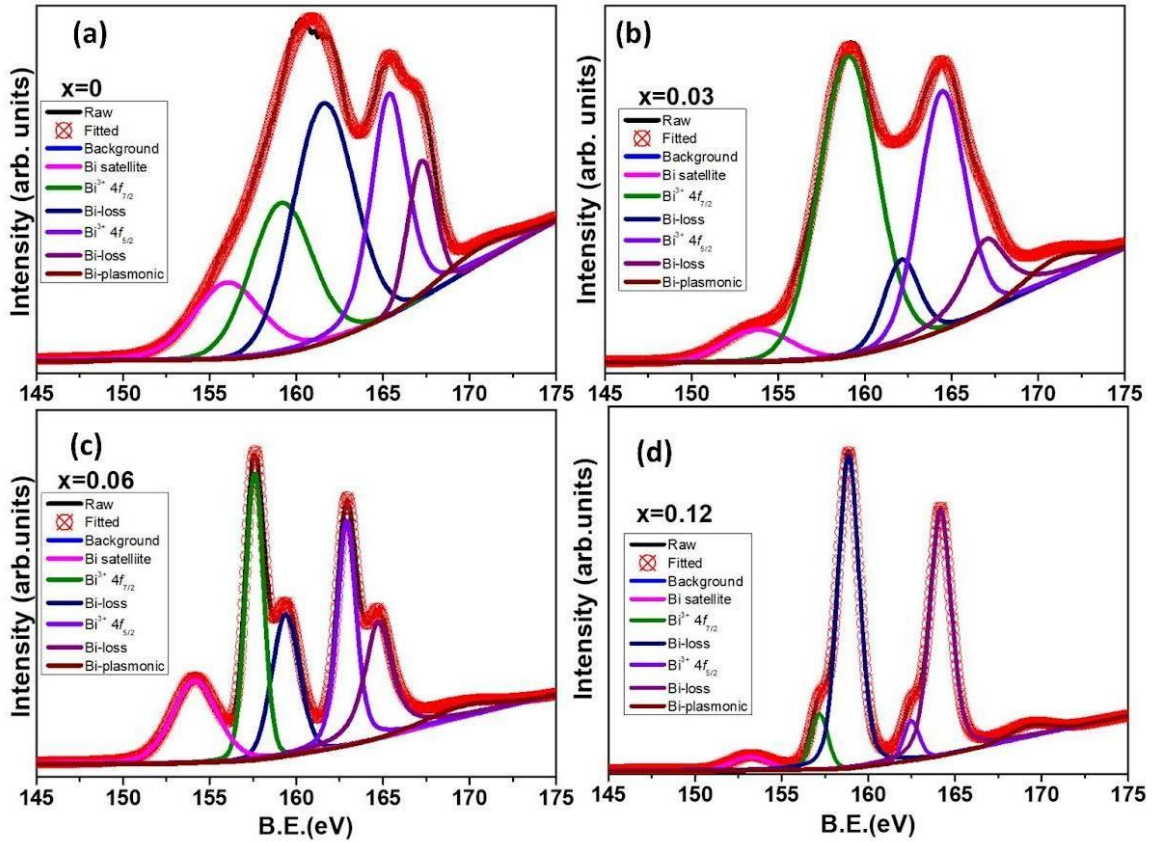

**Figure 5** Deconvoluted Bi-4f XPS pattern of (a) x=0, (b) x=0.03, (c) x=0.06, and (d) x=0.12 compositions

iv. Ti-2$p$ core-level spectra:

The Ti ion is supposed to be in the $Ti^{4+}$ state. However, for all the compositions a presence of a mixed oxidation state of $Ti^{3+}$ and $Ti^{4+}$ was observed [Fig.6(a)-(d)]. The $Ti^{3+}$ $2p_{3/2}$ and $2p_{1/2}$ peaks were observed to increase continuously from (456.68 eV, 462.18 eV) for x=0 to (458.68 eV, 462.77 eV) for x=0.06 but decreased to (457.50 eV, 462.99 eV) for x=0.12 compositions [39]. The same variation was also observed for the average B-O bond strength i.e., the B-O bond strength increased up to x=0.06 and further decreased for the x=0.12. The $Ti^{4+}2p_{3/2}$ and $Ti^{4+}$ $2p_{1/2}$ peaks were observed to continuously increased from (458.26 eV, 463.92 eV) for x=0 to (459.64 eV, 464.37 eV) for x=0.06 but decreased to (458.97 eV, 463.87 eV) for x=0.12 [40]. The spin-orbit splitting for $Ti^{3+}$ states and $Ti^{4+}$ states (5.5 eV, 5.66 eV for x=0) was observed to decreased continuously to (4.09 eV, 4.83 eV for x=0.0625 and thereafter increased to (5.49 eV, 4.90 eV) for the x=0.12 sample. A plasmon peak of Ti is observed in all the samples at ~468.98 eV. Such variations in binding energy affected the <O-B-O> bond angles. The fraction of $Ti^{3+}$ to $Ti^{4+}$ fraction was found to increase continuously from 0.49 for x=0 to 0.53 for x=0.03, 0.58 for x=0.06 to 0.70 for x=0.12 composition respectively. Such preference for $Ti^{3+}$ is due to the substitution of Mn at the Ti site.

The Ti-LMM Auger peaks observed with the Na-1$s$ prominent peak were observed to decrease from 1074.72 eV for x=0 to 1072.07 eV for x=0.12 compositions. This may be due to the increased contribution of Mn. The broad region in the spectra is due to the overlapping of $Bi^{3+}$ $4d_{3/2}$ peak (~465 eV) with $Ti^{4+}$ $2p_{1/2}$ (~463.92 eV) peak, which is observed to be present in all the samples.

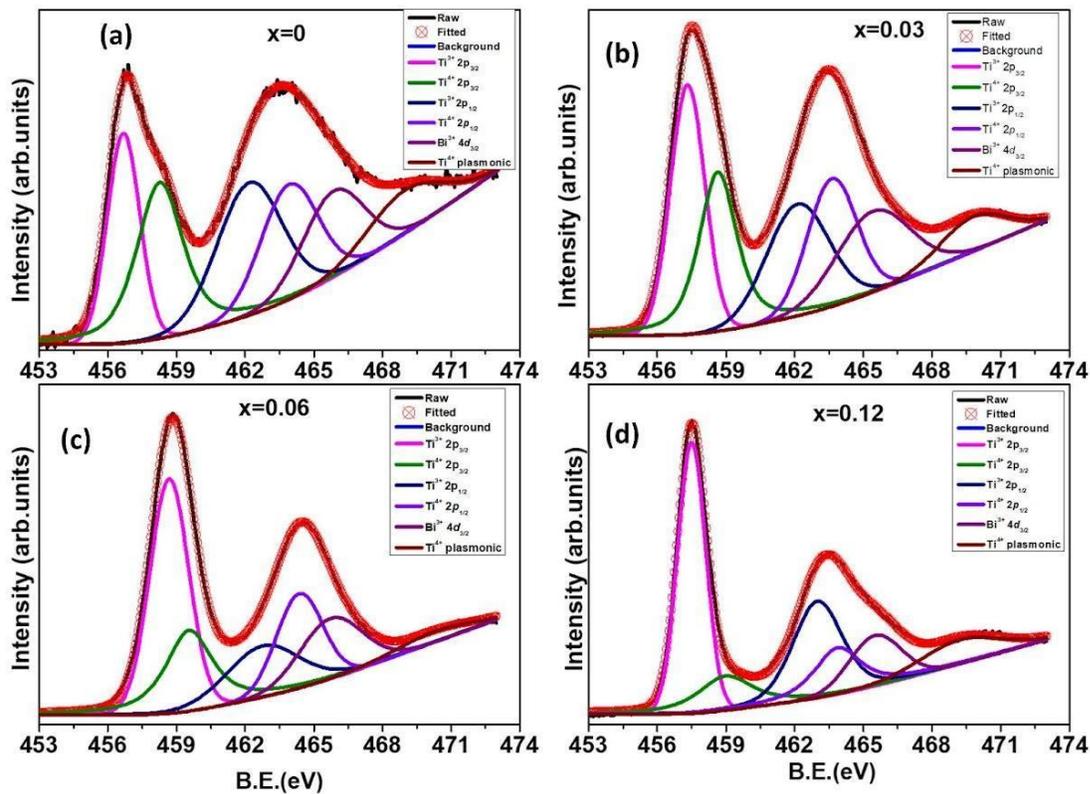

**Figure 6** Deconvoluted Ti-2p XPS pattern of (a) x=0, (b) x=0.03, (c) x=0.06, and (d) x=0.12 compositions

v. Ca-2$p$ core-level spectra:

The Ca-2p core-level spectra confirmed the presence of $Ca^{2+}$ in all the samples [Fig.(a)-(c)]. The $Ca^{2+}$ $2p_{3/2}$ and $Ca^{2+}$ $2p_{1/2}$ was observed to be present at (347.66 eV, 351.58 eV) for x=0.03, (347.67 eV, 351.26 eV) for x=0.06, and (346.36 eV, 349.92 eV) for x=0.12 [41,42].

The B.E. decreased with an increase in Ca content in the sample. The spin-orbit energy was also observed to be continuously decreased from 3.92 eV for x=0.03 to 3.56 eV for x=0.12. Such variation is also reflected in the change in A-O bond strength with composition.

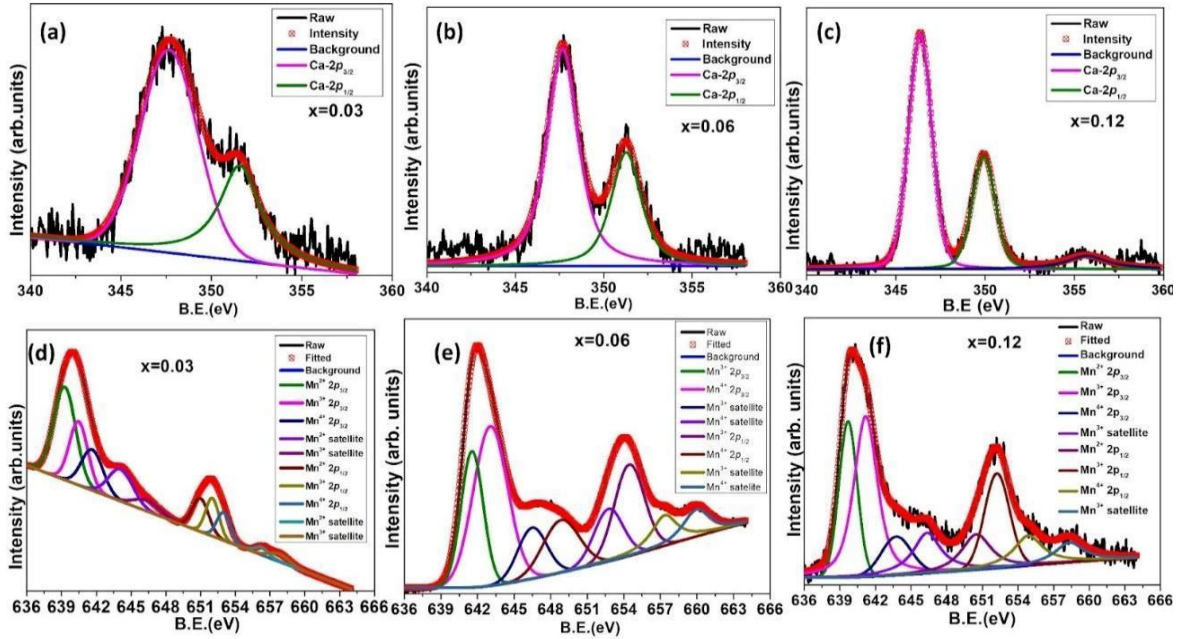

**Figure 7** Deconvoluted Ca-2p XPS pattern of (a) x=0.03, (b) x=0.06, and (c) x=0.12 compositions, Deconvoluted Mn-2p XPS pattern of (d) x=0.03, (e) x=0.06, and (f) x=0.12 compositions

v. Mn 2p core-level spectra:

The Mn-2p XPS spectra was deconvoluted to eight peaks associated with $2p_{3/2}$, $2p_{1/2}$ peaks of $Mn^{2+}$ (639.36eV, 651.24 eV), $Mn^{3+}$ (640.79 eV, 652.54 eV), and $Mn^{4+}$ (642.63 eV, 656.22 eV) and satellite peaks at (644.79 eV, 657.63 eV) corresponding to $Mn^{2+}$, (646.49 eV, 657.34 eV) to $Mn^{3+}$, and (648.85 eV, 659.96 eV) to $Mn^{4+}$ [Fig. 6(a)] [35,39,43,44]. It was observed that there is a presence of a mixed oxidation state of $Mn^{2+}$, $Mn^{3+}$, and $Mn^{4+}$ in the x=0.03 and 0.12 compositions, while only $Mn^{3+}$ and $Mn^{4+}$ states are present in the x=0.06 composition [Fig.7 (d)-(f)]. The $Mn^{3+}$ content was estimated by using the area ratio of Mn3+ to the sum of the area of $Mn^{2+}$, $Mn^{3+}$, and $Mn^{4+}$. It was found that the $Mn^{3+}$ content is nearly the same for both the x=0.03 (0.30) and x=0.06 (0.35) composition while it was increased to 0.51 for the x=0.12 composition. The spin-orbit splitting energy for $Mn^{3+}$ and $Mn^{4+}$ was found to be decreased continuously from (11.53 eV, 11.34 eV) for x=0.03 to (11.20 eV, 11.38 eV) for x=0.06, and (11.05 eV, 11.11 eV) for x=0.125 compositions respectively. The total content of the Mn at Ti place was estimated by multiplying the charge. The corresponding stoichiometry was found to be increased from 0.0868 for x=0.03 to 0.2281 for x=0.06 and 0.3587 x=0.12 compositions. The total charge of the Mn was calculated from the area ratio. It was observed to increase from 2.61 for x=0.03 to 3.65 for x=0.06 and decrease to 2.87 for x=0.12. Such variation in the charge content was reflected in the B-O bond strength variation.

Raman Spectroscopy:

The room temperature Raman spectra were recorded in 100-900 cm$^{-1}$ for all the compositions [Fig.8 (a), (b)]. Twelve Raman active modes are present in the NBT (*R3c*) structure [25]. The first mode A (1) at ~135cm$^{-1}$ continuously redshifted with an increase in the Ca and Mn substitution [Fig. 8(c)]. The FWHM increased while the intensity decreased with Ca and Mn substitution increase. This mode is associated with the vibration of A-site atoms. Such redshift of Raman mode agrees with the decrease in the binding energy of Na-1*s*.

The second mode, E (1) at ~151cm$^{-1}$, continuously blueshifted with an increase in the Ca and Mn substitution [Fig.8(c)]. The mass of Ca is less than the mass of Bi, which decreased the total mass, and hence the blueshift was observed. The FWHM of this mode decreased for x=0.03 and further increased continuously for x=0.06 and x=0.12. This could be associated with the variation of $A_3$-O and $A_4$-O bond lengths.

The mode A (2) and E (2) at ~239cm$^{-1}$ and ~283cm$^{-1}$ continuously redshifts with an increase in the substitution. These modes are due to the complex vibrations of the $BO_6$ octahedra involving angular twisting of the O-cage and central motion of the B-site atoms [25]. The FWHM of these modes increased for x=0.03, decreased for x=0.06, and increased for x=0.12 compositions. Such variations agree with the interpretations of the bond length of the long B-O bond.

Mode A (3) at ~331cm$^{-1}$ redshifted continuously with an increase in substitution [Fig. 8(c)]. These modes are due to a different twisting of $BO_6$ octahedra, respectively. The dominance of $Ti^{3+}$ with an increase in substitution is notable. $Ti^{3+}$ is bigger than $Ti^{4+}$. Hence the B-O bond length increases, thereby weakening the bond strength. This could be the probable reason for the redshift observed in this mode. The FWHM of these modes decreased for x=0.03, increased for x=0.06, and further decreased for the x=0.12 composition. Such variation could be associated with modifying the short B-O bond length and long O-B-O bond angles.

The modes E (3) and E (4) at 526 cm$^{-1}$ and 574cm$^{-1}$ are associated with modifying B-O bonds and the horizontal compression of BO$_6$ octahedra. These two phonon modes vary similarly. A continuous redshift for x=0.03 and x=0.06 and a blueshift for x=0.12 is observed. The FWHM of the E (3) mode increased for x=0.03 and decreased for x=0.06 while again increasing for the x=0.12 composition. An exactly opposite trend was observed for the E (4) mode. Such coexistence of compressive and tensile strain is due to a similar modification in the B-O bond lengths and O-B-O bond angles.

The three modes centered at ~765 cm$^{-1}$, ~828cm$^{-1}$, and ~866cm$^{-1}$ present in the 700-900 cm$^{-1}$ range show the Raman shift similarly [Fig.8(d)]. The Raman mode was redshifted for x=0.03 and blueshifted for x=0.06, and redshifted for the x=0.12 composition. Such variation is associated with the modification in the long B-O bond strength.

Some modes are present that are not theoretically predicted but present in all the NBT-based samples. A Raman mode present at ~479 cm$^{-1}$ is redshifted for x=0.03 and x-0.06 and blue-shifted for x=0.12. The mode at ~610 cm$^{-1}$ blue-shifted for x=0.03 and further red-shifted for x=0.06 and again blue-shifted for the x=0.12 composition. Most probably, these are correlated to the modifications of the B-O bond strengths as the trends are similar. However, this is not a strong claim in the absence of a definite theoretical estimate.

In the pure NBT sample, the Raman modes are strongest in the 200-400 cm$^{-1}$ range with very high intensity while moderate intensities are observed for the modes in the broad range in 400-700 cm$^{-1}$. But with the substitution of CaMnO$_3$, the intensity of these two ranges become equally strong and are comparable. The modes in the ~200-400 cm$^{-1}$ range became broader, while the modes in the range ~400-700cm$^{-1}$ became narrower with an increase in CaMnO$_3$ content. An extra mode was observed at ~412 cm$^{-1}$ in the x=0.06 and x=0.12 compositions. This is a B$_{2g}$ (3) mode associated with the stretching of the MnO$_6$ octahedra of orthorhombic (*Pnm*a) of CaMnO$_3$ [45]. A B$_{1g}$ (5) mode due to the vibration of O-atoms of orthorhombic (*Pnma*) CaMnO$_3$ was observed at 351 cm$^{-1}$ only in the x=0.12 composition [45].

Hence the dominance of the CaMnO$_3$ modes with the increasing substitution is observed highlighting the strong influence of the MnO$_6$ octahedral properties over the TiO$_6$ octahedra in these substituted materials.

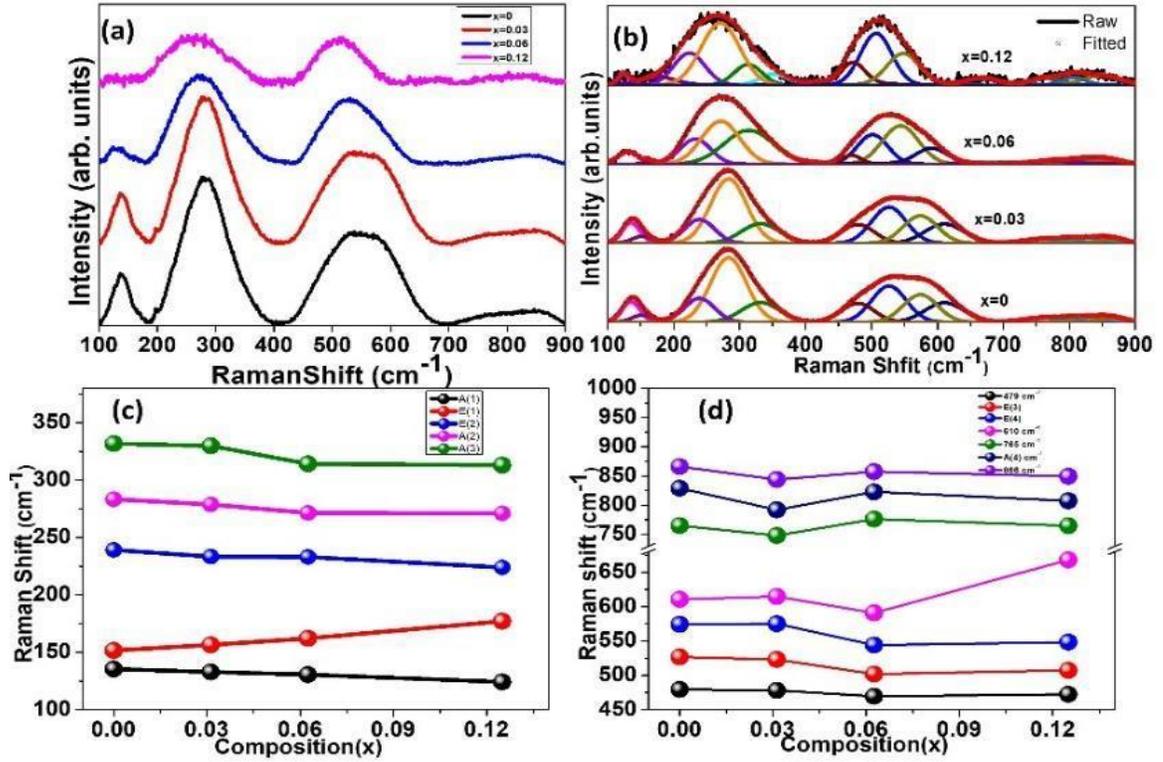

**Figure 8** (a) Raman Spectra for x=0, 0.03, 0.06, and 0.12 compositions, (b) Deconvoluted Raman spectra for x=0, 0.03, 0.06, and 0.12 compositions, (c) and (d) Variation of Raman modes with composition

Morphology Study:

The morphology of all the compositions was studied and analyzed using ImageJ software. All the compositions (x=0, 0.03, 0.06) show a dense honeycomb type morphology [Fig.9(a)-(c)]. But, for the x=0.12 composition, the edges of each grain became rounded rather than a sharp edge [Fig.9(d)]. An additional agglomerated morphology was also observed for the x=0.12 composition, which could be due to two phases in this composition [46]. The x=0.03 composition shows the largest grain size of ~37.78 ±7.68 µm in comparison to the x=0 (15.33 ±4.61µm) composition. The other two compositions, i.e., x=0.06 (8.40 ±2.89 µm) and x=0.12 (13.35 ±6.35µm), show smaller grain size in comparison to the x=0.

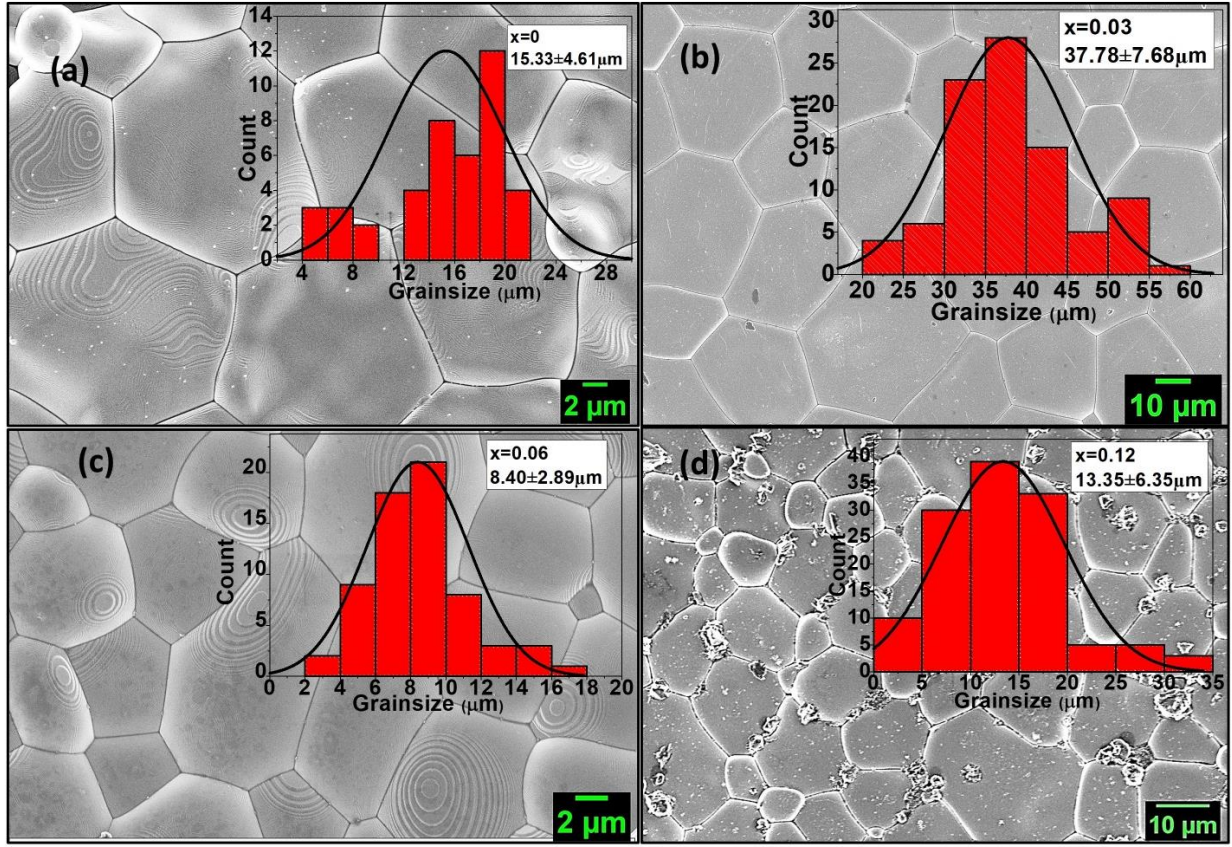

**Figure 9** Surface morphology of sintered pellets for (a) x=0 (b) x=0.03 (c) x=0.06 (d) x=0.12 compositions and the inset show the grainsize distribution curves for the respective compositions

Dielectric Study:

The room temperature dielectric spectra revealed an enhancement in the dielectric permittivity with an increase in the CaMnO$_3$ composition [Fig.10(a)]. The dielectric constant increased for x=0.03 and further continuously decreased for the x=0.06 and x=0.12 compositions. The tan loss also decreased for the x=0.03 composition but continuously increased for the other two. The Bi loss decreased for the x=0.03 and continuously increased for the other two compositions. Hence, such a behavior can be correlated to the Bi loss due to the volatile nature of Bi. This is discussed in the XPS section. As dielectric behavior is dependent upon the grain size of the ceramics, hence the x=0.03 shows the highest room temperature dielectric constant due to its largest grain size [47].

The phase transitions were studied using the temperature-dependent dielectric plot at different frequencies. The phase transition temperature (Tc) corresponding to the ferroelectric to paraelectric phase was observed in the 50 - 400 °C temperature range for the x=0, x=0.03, and x=0.06 compositions [Fig. 10 (b), (c)] [16]. The Tc varies nominally with frequency showing a possibility of a relaxor mechanism. At a frequency of 10kHz, the Tc continuously shifted to a lower temperature with increasing doping, from 332 °C for the x=0 to 275 °C for x=0.03 and to 254 °C for the x=0.062 composition. The phase transition was not detected for the x=0.12 composition [Fig.10 (d)]. This may be due to the near room temperature phase transition of the $CaMnO_3$ phase [48]. For the x=0.12 sample, the $T_c$ may be close to the room temperature. As these measurements were performed above 50 °C, such a transition was not observed. Note that the XRD and Raman analysis of the x=0.12 composition reveals the presence of a mixed phase of *Pnma* and *R3c* phases. While the *R3c* phase shows spontaneous polarization, the *Pnma* contains a centrosymmetric point group "mmm" for which it is paraelectric. Hence, the phase transition temperature was reduced for the x=0.12 near the room temperature.

The NBT is a well-known relaxor-type material. The relaxor nature increased with increased $CaMnO_3$ content. This is due to the increase in diffuseness of the material that may have originated from the random distribution of multiple cations at different atomic sites [16]. The dielectric stability improved significantly over a broad temperature range with $CaMnO_3$ substitution, implying the applicability of these materials as stable capacitors [49]. The $\varepsilon_r/\varepsilon_{rmax}$ vs. $T/T_{max}$ depicts the relaxor nature and stability of the prepared compositions [Fig. 10(f)].

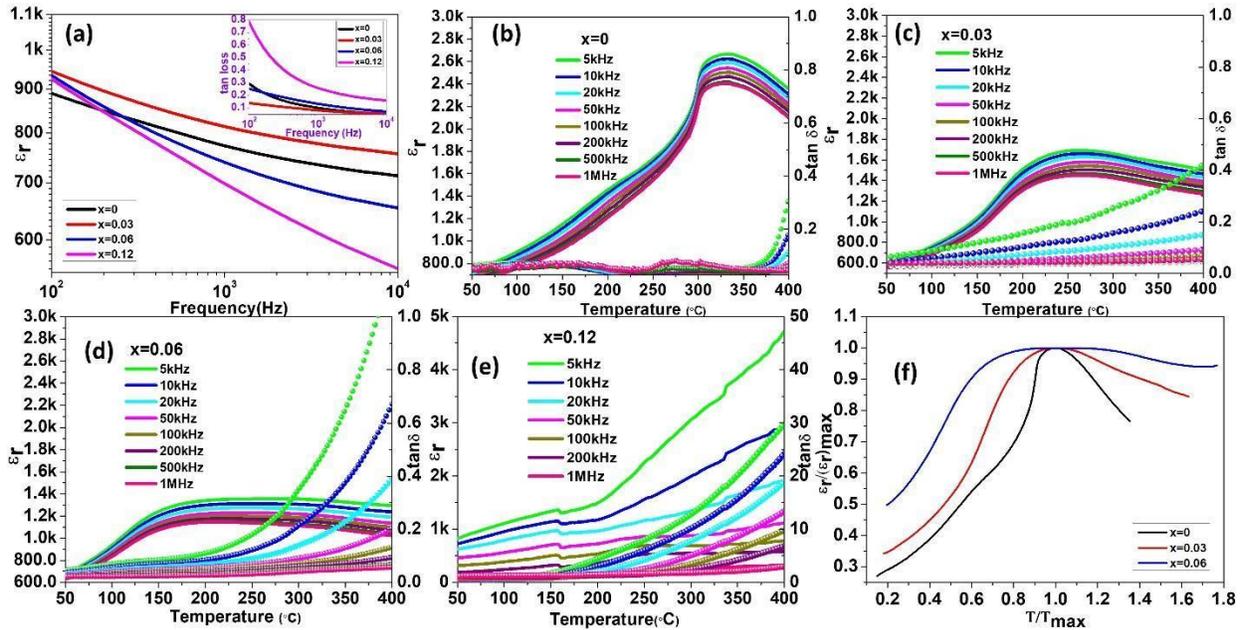

**Figure 10** (a) Comparison of Room temperature dielectric constant and loss for x=0, 0.03, 0.06, and 0.12 compositions, (b), (c), and (d) Temperature dependent dielectric constant and loss for x=0, 0.03,0.06, and 0.12 compositions respectively (f) Variation of Raman modes with composition

Magnetism:

The M–H plot of x=0 ($Na_{0.5}Bi_{0.5}TiO_3$) showed an anti-S-shaped curve which results from the diamagnetic and weak ferromagnetic signals [Fig.11(a)]. It is well known that the diamagnetism of pure $Na_{0.5}Bi_{0.5}TiO_3$ originates from the empty state of $Ti^{4+}$ cations. In contrast, the weak ferromagnetism may create by vacancies (e.g., $Ti^{4-\delta}$ and $Na^+$) or be related to the surface effect [50] [14]. Zhang et al. predicted that perfect $Na_{0.5}Bi_{0.5}TiO_3$ is non-magnetic, while Na or Ti vacancies could induce magnetism rather than the Bi or O vacancies [51]. Ju et al. reported that Na vacancies at the surface were possibly introduced to a non-zero magnetic moment [52]. It is also discussed that the magnetic hysteresis for a Na deficient NBT system showed two distinct behaviors: weak bulk ferromagnetism at low fields (below 800 Oe) and diamagnetism at areas above 800 Oe [17]. In the present work, the XPS revealed the presence of ~49% of $Ti^{3+}$ in the B-site. Also, $Ca^{2+}$ at the $Na^+$ site creates Na-deficiency in the lattice [24]. Such self-defects induced weak ferromagnetism in the pure NBT system.

The M-H plots indicated that the solid solution of $CaMnO_{3-\delta}$ in the host $Na_{0.5}Bi_{0.5}TiO_3$ materials produced a complex magnetic behavior [Fig.11(b-d)]. Some reports suggest that defects at the $Bi^{3+}$, $Na^+$, and $Ti^{4+}$ sites produce a non-zero magnetic moment while oxygen vacancies produce a zero magnetic moment. The reduction of $Ti^{4+}$ to $Ti^{3+/2+}$ induces magnetic moment in the NBT-based systems. The incorporation of $CaMnO_3$ reduced the diamagnetism component in the modified systems. The magnetic moment is relatively the same ~0.028emu/g for x=0.03 to ~0.03emu/g for x=0.06 and further improved to ~0.10 emu/g for x=0.12 compositions at 30kOe field. The M-H plots show an unsaturation of magnetization with an applied magnetic field, possibly due to the competing effect among the ferromagnetic, paramagnetic, and antiferromagnetic components that give rise to the total magnetic moment of all the compositions[24].

The x=0.03 composition shows a mixed behavior of paramagnetism and very weak ferromagnetism [Fig.11(b)]. Such existence of weak ferromagnetism could be due to a handful of different possible interactions. From the XPS analysis, it was observed that there is a presence of $Mn^{2+}$, $Mn^{3+}$, and $Mn^{4+}$ in this composition. The amount of $Mn^{2+}$ is ~50% while $Mn^{3+}$ is ~35% and $Mn^{4+}$ is ~15%.

The interaction of $Mn^{2+/3+}$ through oxygen vacancies ($V_O$) called the F-centre exchange interaction induced the ferromagnetism in this system [53]. This ferromagnetic ordering is due to several favorable $Mn^{2+/3+}$+- ($V_O$) -$Mn^{2+/3+}$. With the substitution of $Ca^{2+}$ for $Na^+$ at the *A*-site, charge compensation demands the introduction of Na-vacancies. These vacancies influenced ferromagnetism. Also, the $Ti^{3+}$ defects induce ferromagnetism. As Ti3+ is ~53% in this system calculated from the XPS analysis, some ferromagnetism could be due to the $Ti^{3+}$ states. The presence of paramagnetism was possibly related to the isolation of Mn cations, which favored the paramagnetic property in this material [54].

The x=0.06 composition shows a mixed behavior of paramagnetism and weak antiferromagnetism [Fig.11 (c)]. Generally, the bulk $CaMnO_3$ is known to show G-type antiferromagnetism. $Mn^{3+}$ has a configuration of $t_{2g}^3$ $e_g^1$, while $Mn^{4+}$ has a $t_{2g}^3$ configuration. G-type antiferromagnetism originates from the hopping of $e_g$ electron from $Mn^{3+}$ to $O^{2-}$ and then from $O^{2-}$ to $Mn^{4+}$. In the case of G-type antiferromagnetism, both the interplanar and intraplanar coupling are antiferromagnetic [22,55]. From the XPS studies, it is revealed that this composition contains $Mn^{3+}$ and $Mn^{4+}$ where the $Mn^{3+}$ is ~35% while $Mn^{4+}$ is ~65%. However, the $Mn^{4+}$ cation incorporated for the $Ti^{4+}$- site resulted in the antiferromagnetic interaction of the $Mn^{4+}$- $O^{2-}$–$Mn^{4+}$ pair creating the superexchange interaction [56]. Also, this composition contains ~60% $Ti^{3+}$, which could result from the interaction of $Ti^{3+}O^{2-}$-$Ti^{3+}$, hence the antiferromagnetic property.

The x=0.12 composition shows a weak ferromagnetic behavior [Fig.11 (d)]. The XPS study reveals the presence of $Mn^{2+}$ (~31%), $Mn^{3+}$ (~51%), and $Mn^{4+}$ (~18%) in this composition. Also, the $Ti^{3+}$ is ~70% in this composition. The increase in $Ti^{3+}$ is due to the change of $Ti^{4+}$ cations due to the influence of oxygen vacancies which also contributes to the source ferromagnetism because of the conversion of an empty 3d shell to an occupied $3d$ shell in $Ti^{3+}$ cations [14]. Here, the oxygen vacancy enhanced the F-centre exchange interaction between the $Mn^{2+/3+}$- $V_O$ -$Mn^{2+/3+}$. There is also some presence of antiferromagnetism in all the $CaMnO_{3-\delta}$ modified compositions. In addition to the above-discussed possibilities, such behavior could also be due to the interaction between the polaron of $Mn^{2+/3+}$- $V_O$ - $Mn^{2+/3+}$ that arises from the non-uniform incorporation of Mn ions into the parent lattice [24].

The ferromagnetism behavior in the samples is represented in Fig. 12 (a) and (b). The other magnetic contributions, like paramagnetic, diamagnetic, and antiferromagnetic, were subtracted from the M(H) loop to extract the ferromagnetic contributions for all the compositions. The remnant magnetization and the saturation magnetization were calculated from the ferromagnetic loop to understand its variation with the composition. The remnant magnetization ($M_r$) was observed to be decreased continuously up to the x=0.06 composition while it increased for the x=0.12 composition, and a similar variation was also noted for the saturation magnetization ($M_s$) as well [Fig. 12 (c)]. In addition to the above discussed ferromagnetic exchange interactions, the $Ti^{3+/4+}$ and $Mn^{3+/4+}$ double exchange interactions mediated via the $O^{2-}$ ion ($Ti^{3+}$-$O^{2-}$-$Mn^{3+}$, $Ti^{3+}$-$O^{2-}$-$Mn^{4+}$, $Ti^{4+}$-$O^{2-}$-$Mn^{3+}$, $Ti^{4+}$-$O^{2-}$-$Mn^{4+}$) also contribute to the ferromagnetism in $CaMnO_3$ modified samples [57,58].

The B-O-B bond angle plays a vital role in deciding the type of magnetism in a material [59]. In this series of compositions, the x=0 to x=0.06 composition shows an *R3c* structure while the x=0.12 shows a mixed phase of rhombohedral *R3c* and orthorhombic *Pnma* structure. In the *R3c* structure, the B-O-B bond angle continuously increased from 157.70 for x=0 to 164.08 for x=0.06 and decreased to 157.21 for the x=0.12 composition. For $d^3$ $Mn^{4+}$ cations in octahedral coordination, G-type antiferromagnetic superexchange is favored by 180° transition metal-oxygen–transition metal bond angles and is weakened as the B-O-B angle deviates from this angle [60,61]. The deviation from this angle continuously decreased from x=0 to x=0.06 (lowest deviation) and increased for the x=0.12 composition. This indicates the introduction of antiferromagnetism in the x=0.03 and x=0.06 while the antiferromagnetism is highest in the x=0.06 composition.

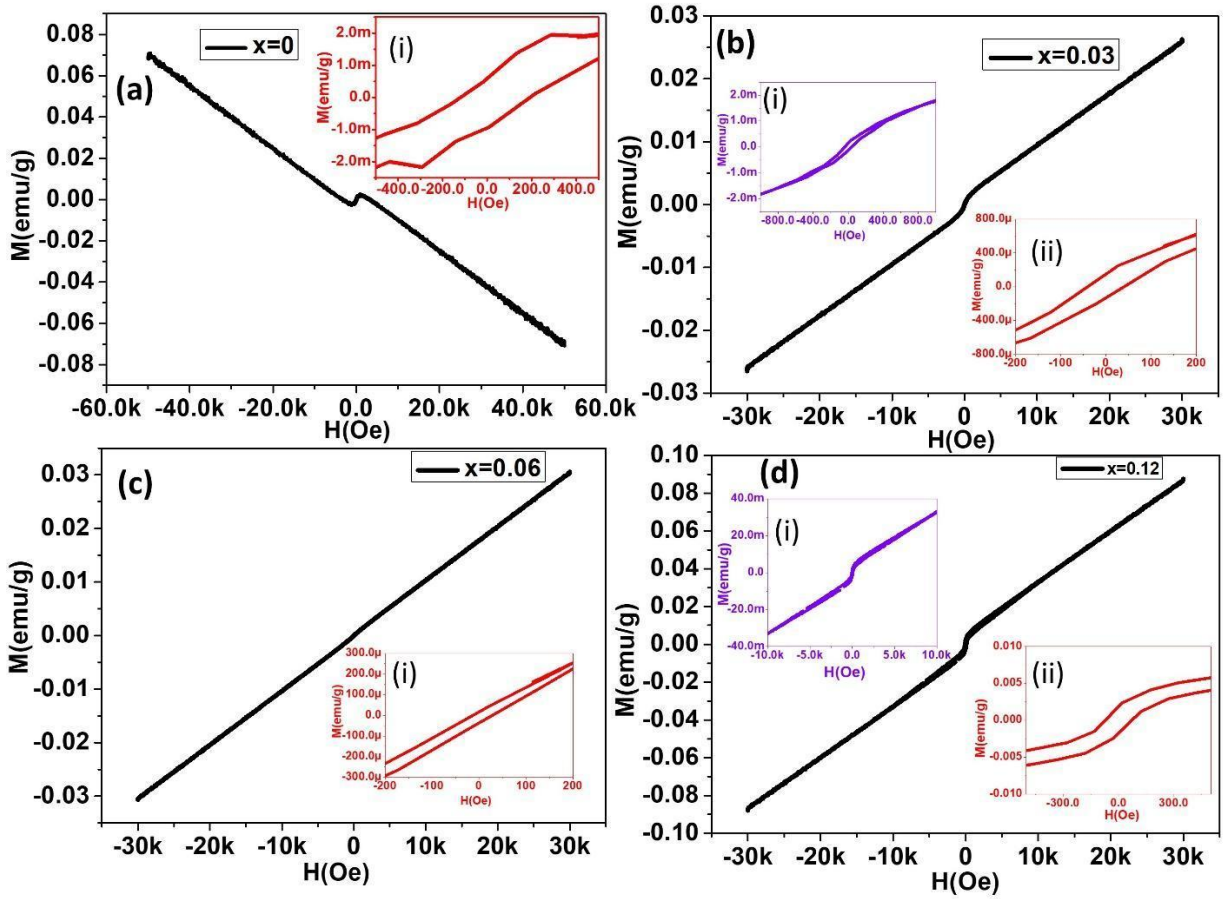

**Figure 11** Room temperature M-H curve for (a) x=0 (b) x= 0.03 (c) x= 0.06, and (d) x= 0.12 compositions; The inset (i) and (ii) shows the magnified region confirming a ferromagnetic behaviour in all the samples

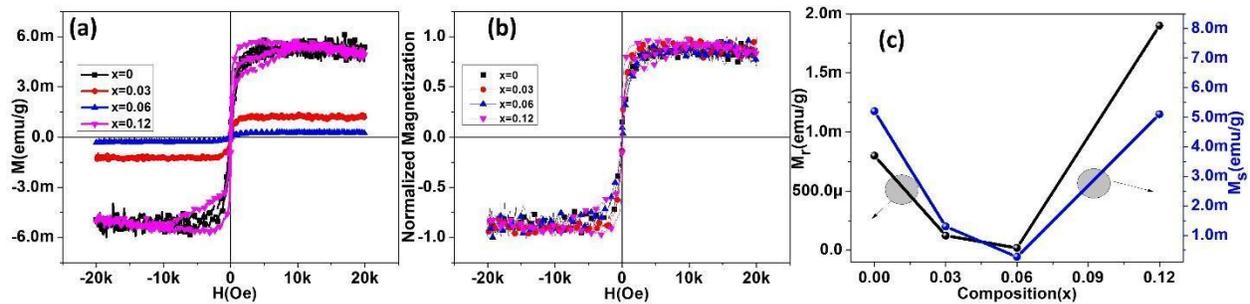

**Figure 12** (a) Room temperature ferromagnetism plots after extracting the diamagnetic, paramagnetic, and antiferromagnetic contributions for x=0, 0.03, 0.06, and 0.12 compositions (b) Normalized ferromagnetic contributions for x=0, 0.03, 0.06, and 0.12 compositions (c) Variation of remnant magnetization and saturation magnetization with composition

Magnetodielectric Coupling:

To observe the effect of the magnetic field on the dielectric properties of each composition, the dielectric constant was measured under different applied magnetic fields [Fig.13]. The tuning of polarization in a magnetic field is governed by magnetodielectric coupling, also called magnetoelectric (ME) coupling [6,17]. It is well known that the ME coupling can be examined from the magneto-dielectric measurement [8,62]. The change in the dielectric permittivity can validate the ME coupling in a material by applying an external magnetic field [62]. To investigate the room temperature magneto-dielectric effect in the $CaMnO_3$ modified NBT compositions, a frequency-dependent dielectric measurement was done at various magnetic fields at room temperature. The capacitance spectra were observed to be varied with an increase in the applied magnetic field for all the compositions. For the x=0, 0.06, and 0.12 compositions, the Cp was observed to decrease with applying a magnetic field consistently. This proves the coupling between the magnetic field and the dielectric constant of the respective materials. An irregular behavior was observed in the x=0.03 composition, which is a decrease in the capacitance at a lower magnetic field of 2kGauss and further increased continuously. To understand the coupling in detail, the magnetodielectric constant MD was calculated using the following equation: MD%=[($C_P$(H)-$C_P$(0))/$C_P$] ×100, where $H$ and 0 stand for the dielectric constants under applied magnetic field and zero applied magnetic field, respectively [12]. MD% is an important parameter to compare the ME coupling in a material [7,52]. It can be observed that with the increase in applied magnetic fields, the MD% value decreased for the x=0, x=0.06, and x=0.12 compositions [Fig.14 (a), (c), and (d)]. But for the x=0.03 composition, it initially decreased to 2kGauss and further increased for the higher applied fields [Fig.14 (b)]. A similar nature was observed for all the frequencies from 1kHz to 1 MHz. The MD% increased with the increase in the frequency. The higher value of MD% at low frequency may be contributed by dc conductivity, leakage current, space charge polarization, etc. [62]. In the high-frequency region around 100 kHz, which is considered the intrinsic region, MD% at 5kGauss was observed to be highest (-3.69%) for the x=0.06 composition. All the compositions show a negative ME coupling, while the x=0.03 composition showed a mixed negative and positive coupling. This could be associated with the competing effect between the lattice strain and the observed mixed magnetism of paramagnetic, antiferromagnetic, and ferromagnetic in this material. The ME coupling increased with an increase in the $CaMnO_3$ incorporation, while a sudden drop in the coupling percentage was observed for the x=0.12 composition. As the x=0.12 composition show the presence of mixed phase of rhombohedral (*R3c*) and orthorhombic (*Pnma*), the ferroelectric polarization is less in this material due to which less coupling was observed in comparison to the other compositions as ME coupling is best observed in the multiferroic materials [5,63]. In all the compositions, a hysteresis was observed for the MD% with an upward field and withdrawal of the field.

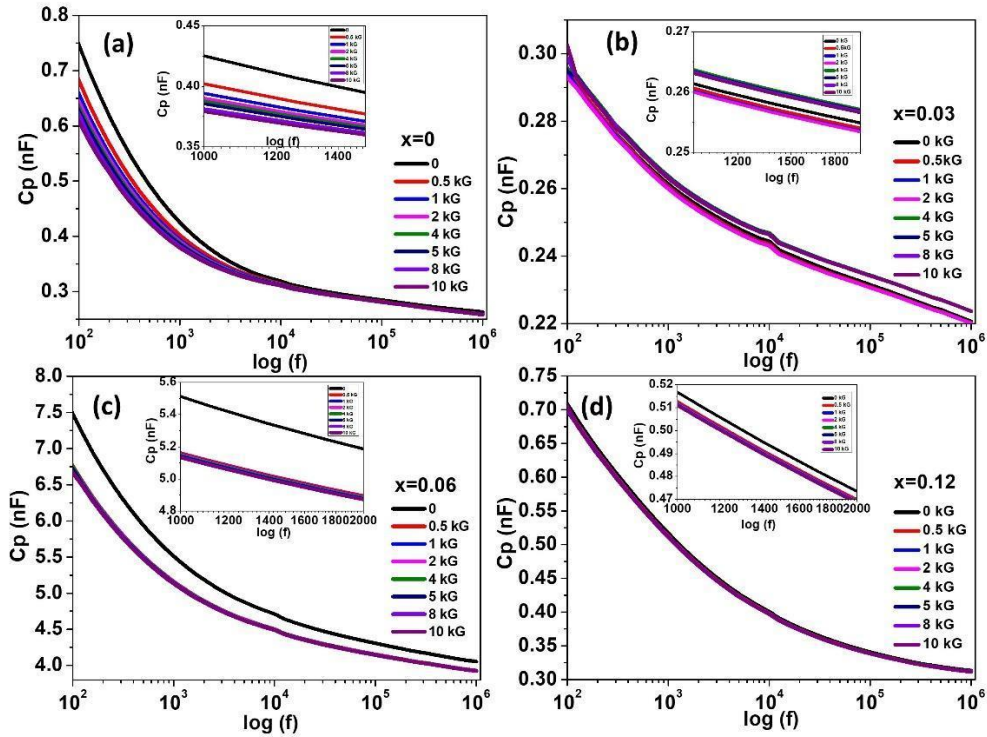

**Figure 13** Variation of Capacitance with frequency from 100 Hz to 1MHz at different applied magnetic fields for (a) x=0 (b) x= 0.03 (c) x= 0.06, and (d) x= 0.12 compositions; the inset shows a zoomed picture of the same

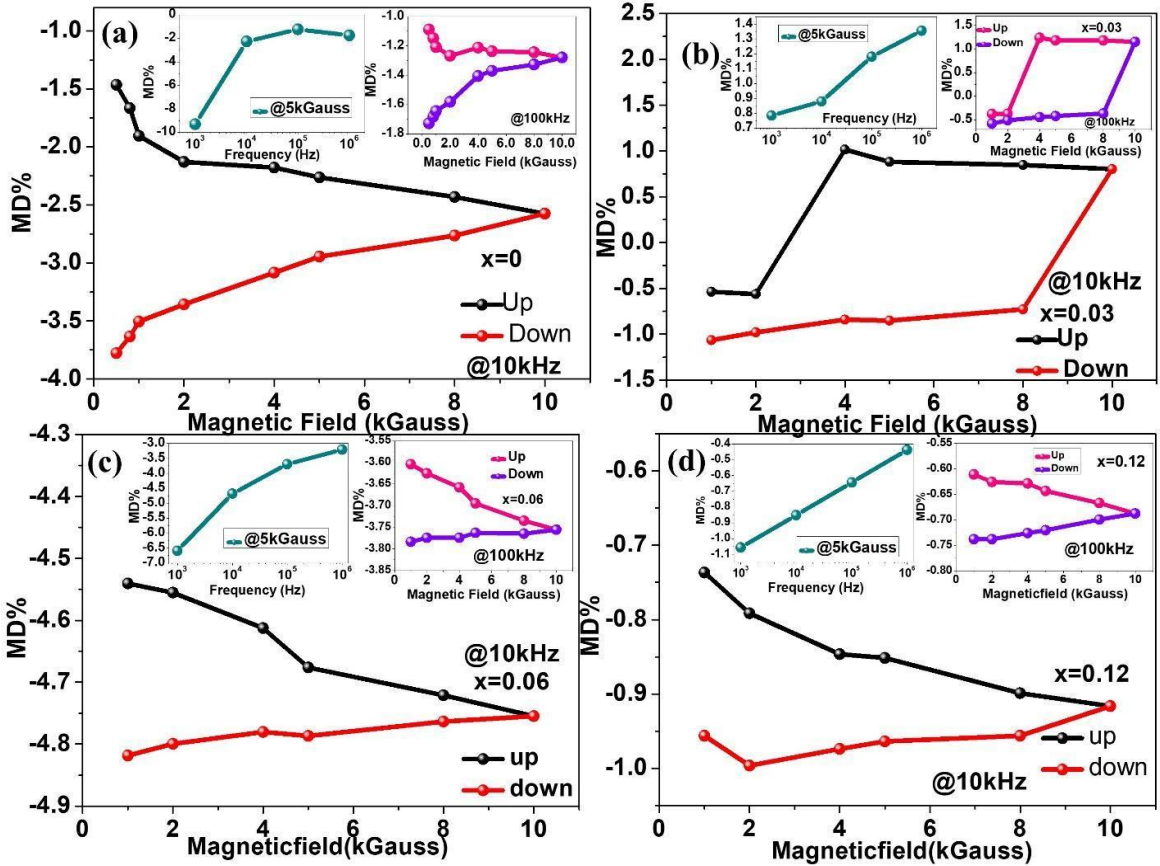

**Figure 14** Variation of MD% with an applied magnetic field at Room temperature for (a) x=0 (b) x= 0.03 (c) x= 0.06, and (d) x= 0.12 compositions; the inset shows variation of MD% with an applied field at 100kHz and the variation of MD% with frequency.

**Conclusion:**

The prepared (1-x) $Na_{0.5}Bi_{0.5}TiO_3$-xCaMnO$_3$ compositions show a Rhombohedral (*R3c*) phase for x≤0.06 while a mixed Rhombohedral (*R3c*) and orthorhombic (*Pnma*) phase for the x=0.12. The rhombohedral cell volume was observed to be decreased with CMO incorporation and while lattice strain decreased for x=0.03 and further increased for the other compositions. The valence state study revealed such variation in the strain is dominated by the Bi loss in the respective compositions. The bond lengths corroborated the Raman shift of all the modes for all the compositions and bond angles observed from the structural analysis. The $B_{2g}$ (3) mode at ~412 cm$^{-1}$ originated due to the stretching of MnO$_6$ octahedra of orthorhombic (*Pnm*a) observed for both the x=0.06 and 0.12 compositions. Another mode called $B_{1g}$ (5) mode 351 cm$^{-1}$ due to the vibration of O-atoms of orthorhombic (*Pnma*) CaMnO$_3$ is only present in the x=0.12 composition, confirming the presence of mixed-phase. The grainsize was observed to be highest for the x=0.03 composition. The room temperature dielectric constant and loss were improved for the x=0.03 composition due to less Bi and O vacancy and the largest grain size among all the compositions. The lowering of the ferroelectric to paraelectric phase transition temperature (Tc) towards the room temperature was achieved by the incorporation of CMO. The magnetism study revealed a presence of weak ferromagnetism in all the compositions due to the presence of various self-defects like Na-vacancy and Ti$^{3+}$ states. The CMO modified compositions contain ferromagnetism due to the self-defects and the different exchange interactions like F-center exchange between Mn$^{2+/3+}$- V$_O$ - Mn$^{2+/3+}$ and double exchange interaction between Ti$^{3+/4+}$ and Mn$^{3+/4+}$ mediated by O$^{2-}$ ion. The magnetization was also improved with the CMO incorporation in the NBT lattice. The coupling between the magnetic and dielectric polarization was observed for all the compositions confirming a magnetodielectric coupling. Hysteresis in the Magnetodielectric coupling was also observed with applying and withdrawing the magnetic field for all the compositions. The highest negative MD% of 3.69% at 100kHz (applied field of 5kGauss) was observed for the x=0.06 composition making it a promising material for many magnetoelectric device applications.